\documentclass{amsart}

\usepackage{graphicx}
\usepackage{amssymb,amsmath,amsxtra}
\usepackage{array}

\begin{document}

\newcommand{\pe}{\psi}
\def\d{\delta}
\def\ds{\displaystyle}
\def\e{{\epsilon}}
\def\eb{\bar{\eta}}
\def\enorm#1{\|#1\|_2}
\def\Fp{F^\prime}
\def\fishpack{{FISHPACK}}
\def\fortran{{FORTRAN}}
\def\gmres{{GMRES}}
\def\gmresm{{\rm GMRES($m$)}}
\def\Kc{{\cal K}}
\def\norm#1{\|#1\|}
\def\wb{{\bar w}}
\def\zb{{\bar z}}

\def\bfE{\mbox{\boldmath$E$}}
\def\bfG{\mbox{\boldmath$G$}}

\title{Set Reduction In Nonlinear Equations}

\author{Erhan Turan}
\address{Department of Mechanical Engineering, Bo\u{g}azi\c{c}i University,
Istanbul, Turkey}
\curraddr{Department of Computer Science,
ETH Z\"{u}rich, Z\"{u}rich, Switzerland}
\email{turan@gmail.com}
\thanks{This work was supported by Bo\u{g}azi\c{c}i University Research Fund Grant BAP 07A601D}

\author{Ali Ecder}
\address{Department of Mechanical Engineering, Bo\u{g}azi\c{c}i University,
Istanbul, Turkey}
\email{ecder@boun.edu.tr}


\date{March 1, 2012.}


\begin{abstract}
In this paper, an idea to solve nonlinear equations is presented. During the solution of any problem with Newton's Method, it might happen that some of the unknowns satisfy the convergence criteria where the others fail. The convergence happens only when all variables reach to the convergence limit. A method to reduce the dimension of the overall system by excluding some of the unknowns that satisfy an intermediate tolerance is introduced. In this approach, a smaller system is solved in less amount of time and already established local solutions are preserved and kept as constants while the other variables that belong to the ``set" will be relaxed. To realize the idea, an algorithm is given that utilizes applications of pointers to reduce and evaluate the sets. Matrix-free Newton-Krylov Techniques are used on a test problem and it is shown that proposed idea improves the overall convergence.
\end{abstract}

\maketitle

\section{Introduction}
Newton's Method (NM) is frequently used to solve systems of nonlinear equations. This method starts with an initial guess, $\mathbf{x}^0$ and after $k$ updates, the iterate $\mathbf{x}^k$  satisfies the Equation \ref{eq:nonlinear}; here $\mathbf{f}$ denotes discretized form of the field equations. To find an update, the linear system stated in Equation \ref{eq:jacobian} should be solved at each step, $\mathbf{J}$ being the Jacobian Matrix such that $J_{ij}=\frac{\partial f_i}{\partial x_j}$. One Newton step is completed as current estimate of the solution vector is updated with Equation \ref{eq:update} where $\lambda$ acts as a damping parameter.
\begin{equation}
\mathbf{f}(\mathbf{x})=0\label{eq:nonlinear}
\end{equation}
\begin{equation}
\mathbf{J}\mathbf{\Delta x}=-\mathbf{f}\label{eq:jacobian}
\end{equation}
\begin{equation}
\mathbf{x}^{k}=\mathbf{x}^{k-1}+\lambda\mathbf{\Delta x}\label{eq:update}
\end{equation}
Main advantage of the method is that the computations converge quadratically if the estimate is close to the exact solution. However, there are some downsides of the method. First is the selection of the initial guess. A good initial guess can lead to convergence in a couple of steps. To improve the initial guess, several ideas can be tried \cite{knoll_2004a}. As a starter, the problem can be solved on a coarser mesh first and then the solution can be interpolated on to the fine grid. One can also perform continuation via the physical parameters of the problem, like the Reynolds number, $Re$, in CFD. Another idea to enhance the estimate is to use pseudo-transient algorithms at which the problem is allowed to advance in time at which the steady state will yield the solution of the discrete problem.

Another issue in NM is the solution of Equation \ref{eq:jacobian}. Two things should be considered. The Jacobian, $\mathbf{J}$ should be formed at each step and proper linear solvers should be applied to calculate the update $\mathbf{\Delta x}$. Formation of the Jacobian is not trivial to accomplish since the matrix is generally large and sparse and analytic representation of matrix is not available. Color-based evaluation techniques \cite{gebremedhin_2005a} or Automatic Differentiation \cite{griewank_2008a} can be used for efficient computation of the Jacobian. Still, storage of the entries is the main concern and appropriate sparse storage schemes should be used \cite{saad_2003a}. When Newton-Krylov Methods are utilized, than the formation of the Jacobian could be avoided (unless preconditioner is needed) and the matrix-vector products can be calculated with directional differencing \cite{brown_1990a}.

One final concern is the global convergence. If the estimate is not close to the exact solution, i.e. when the computation starts from rather an uneducated guess - damping is required to calculate the solution. Using a constant $\lambda$ such that $0<\lambda\leq1$ to damp the update is useful to improve the convergence. On the other hand, selection of the damping parameter should be performed with line search or trust region methods \cite{dennis_1983a} so that full steps can be used as the iterates approach the exact solution.

In the literature, several studies are performed on Newton's Method. This paper focuses on speeding up the computations by reducing the solution set. In the solution domain, it might happen that some points do not satisfy the convergence criteria as fast as the rest because of the local nonlinearities. These problematic nodes lag the convergence, in other words, the method continues until all unknowns satisfy the tolerance. In the meantime, already converged points are kept in the solution set. The idea is to isolate those points with high local residual and solve them alone. Hence the object of this study to analyze the performance of the subsets created while solving Equation \ref{eq:nonlinear}.

This paper is organized as follows. In section \ref{set_idea}, the set idea is presented and various solution strategies are discussed. Section \ref{algorithm} states the algorithm and gives insight about the implementation. In section \ref{results}, numerical techniques used in the study are explained and the performance of the presented idea is demonstrated on a test problem. Last section is devoted for conclusion and remarks for future work.
\section{Set Idea}\label{set_idea}
While solving a nonlinear system, the convergence of some of the unknowns might be different than others. These unknowns might be a member of a different equation or they might be located on challenging parts of the domain. In CFD for instance, continuity equation converges slower compared to momentum equations and yet the computations stop only if the mass conservation is satisfied. From this observation, one is can ask the following question: can the solution set be restricted into a subset such that the solver concentrates just on those gradually converging components? In set idea, the subset it is not restricted with just one equation or parts of the solution domain. All of the unknowns are handled at the same time and a subset should be selected which in turn reduces computational time but converges as before. Such a methodology requires two things, a criteria to decide on the subset and a data structure that can handle the changes on the solution set.

Before proceeding, a few remarks on Newton's Method (NM) should be stated. When NM is tested for convergence, what is generally used as a criteria is the norm of vector, $\left\| \mathbf{f} \right\|_2$. However, this norm is just a representative scalar which shows how well the iterate, $\mathbf{x}^k$ satisfies Equation \ref{eq:nonlinear}. It is possible that for a not converged problem, some of the points are already satisfactory in terms of the convergence criteria and yet they are kept in the system until the solution is found. This observation reveals one thing; the solution might be performed on a smaller set as soon as some of the points satisfy the convergence criteria. Hence, the set idea is merely simple: Select some points on the solution domain with a ``set rule" and perform computations only for the unknowns related to those points while keeping the others fixed i.e. freezed. The intention is isolate the points above the RMS limit and to solve only those points. Normally, it might be useful to use different $\lambda_i$'s for each $x_i$, so the nonlinearity can be controlled locally, however, selection of damping parameters one-by-one is a tedious task and the selection criterion is not obvious. Instead, points with similar behavior can be grouped together and handled separately.

In the proposed idea, a Global set, denoted by $\mathcal{S}_G$ includes all of the unknowns of the problem. This is the set that the answer is looked for. A local set, $\mathcal{S}_L$, on the other hand, is a subset of $\mathcal{S}_G$ which is determined in run-time. An obvious outcome is that $\mathcal{S}_L\subseteq\mathcal{S}_G$. When both sets are equivalent, then Newton's Method performs as usual. If, however, a subset is generated than Equation \ref{eq:jacobian} is solved for a smaller problem.

In \cite{lanzkron_1996a}, a special case of the proposed idea is presented. The so called Nonlinear Elimination is used to eliminate some of the unknowns-just to concentrate only on the harder part of the model. What is meant with hard is that parts of the domain are involved with high nonlinearities with degrades the efficiency of the solution process. To get rid of the difficulties, it is suggested to solve that particular domain with high accuracy while keeping other values fixed. As a next step, the unknowns are solved all together in such a way that the initial guess is improved by keeping the solution on the non-eliminated domain intact. What is missing in this methodology is the automation of the eliminated points. Eliminated nodal values are defined a priori.

To decide on a set, three different methods are suggested to be used as a rule. Residual based criteria is stated in Equation \ref{eq:setC1}. $\alpha \leq 1$ is an additional parameter to control the tolerance. A small $\alpha$ will force a stricter selection criterion. As seen in Equation \label{eq:setC1}, Root-mean-square (RMS) is used to examine local residuals. Use of RMS is more appropriate to perform a comparison with local residuals, $f_i$'s since the nonlinear norm, $\left\| \mathbf{f} \right\|_2$ is the sum of all local norms and it is obviously larger than all of the residuals.
\begin{equation}
if\,\left\{ {\begin{array}{*{20}c}
   {f_i  < \alpha\left\| \mathbf{f}  \right\|_{RMS} ,\,\,\,\,\,f_i  \notin \mathcal{S}_L }  \\
   {f_i  \geq \alpha\left\| \mathbf{f}  \right\|_{RMS} ,\,\,\,\,\,f_i  \in \mathcal{S}_L }  \\
 \end{array} } \right.\label{eq:setC1}
\end{equation}
One other idea might be to compare local residual against the average of local residuals. This idea would be especially useful for problems with more than one local degree-of-freedom (dof). In multi-dof problems, dof can be compared against the averages of that particular dof itself instead of using an average of all unknowns.
\begin{equation}
if\left\{ {\begin{array}{*{20}c}
   {f_i  < \alpha {{\left( {\sum\limits_{j = 1}^N {f_j } } \right)} \mathord{\left/
 {\vphantom {{\left( {\sum\limits_{j = 1}^N {f_j } } \right)} N}} \right.
 \kern-\nulldelimiterspace} N}\,\,\,,\,\,\,f_i  \notin \mathcal{S}_L }  \\
   {f_i  \geq \alpha {{\left( {\sum\limits_{j = 1}^N {f_j } } \right)} \mathord{\left/
 {\vphantom {{\left( {\sum\limits_{j = 1}^N {f_j } } \right)} N}} \right.
 \kern-\nulldelimiterspace} N}\,\,\,,\,\,\,f_i  \in \mathcal{S}_L }  \\
 \end{array} } \right.
\label{eq:setC2}
\end{equation}
Yet another idea is to compare the size of the unknown $x_i$ against the step size $\Delta x_i$. It is natural that the computations should continue if the update is larger than the value that it is applied to. To proceed one step further, one should keep the unknown in the set if the update is larger even if the two previous selection criteria are satisfied.
\begin{equation}
if\left\{ {\begin{array}{*{20}c}
   {\left| {\Delta x_i } \right| < \,\left| {x_i } \right|\,,\,\,\,f_i  \notin \mathcal{S}_L }  \\
   {\,\left| {\Delta x_i } \right| \geq \,\left| {x_i } \right|\,,\,\,\,f_i  \in \mathcal{S}_L }  \\
 \end{array} } \right.\label{eq:setC3}
\end{equation}
Note that first two criteria still work with constant values, i.e. the operation is identical at each point. Last criterion is, on the other hand, unique for each dof.

In numerical tests, it is observed that residual based decision criteria might not improve the solution. The solution completes as $\mathcal{S}_G$ at its best if a strict criterion i.e. a small $\alpha$ is used. A looser criterion ($\alpha=1$) might need more Newton iterations as well as more linear iterations per Newton step. Also, the convergence history is not smooth as the global problem, though for shorter run-times one can sacrifice on that, and yet sample computations also take more time as demonstrated in section \ref{results}.

Since the residual alone does not reflect the actual stand of the nonlinear problem well, the update should also be considered. As a matter of fact, the magnitude of the update should be compared with the solution as in the third rule. This condition is also used in $\mathcal{S}_G$ and essential  to capture the physics \cite{shadid_1997a}.

Major difference with the Nonlinear elimination is that the local set is defined during run time. Also it is not fixed, it can adapt itself as the solution proceeds with subsequent Newton iterations. Another distinction with Nonlinear elimination is that an initial attempt is made to calculate the first update vector $\Delta x$. The decision on $\mathcal{S}_L$ is made after this first step. Hence, first it is worked on global set and decide on the local set. Nonlinear elimination, on the other hand starts with a predetermined local set. Later, this information is used to improve the initial guess on the global set.

When the initial norm is computed, the nonlinear residual is calculated. That means, a guess for the subset could be made with this piece of information. Unfortunately, initial estimate might lack essential information regarding to the physics (if for example a zero initial vector is used, convective terms in Navier-Stokes equations are ignored which are the source of nonlinearity at the first place!). Additionally, it is already suggested that magnitude of step sizes are also useful to control the subset. $\Delta x$, however, is set to be zero as a rule of thumb in Newton's Method so there is no information regarding the magnitudes of local updates beforehand.
\begin{figure}[htbp]
\centering
\includegraphics[scale=0.4]{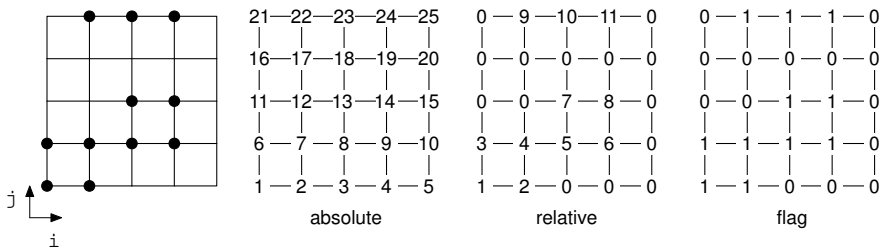}
\caption {Demonstration of the set idea}
\label{f:simpleGridSet}
\end{figure}

Consider Figure \ref{f:simpleGridSet}. This is an example grid to demonstrate the set idea. Assume that the points marked with black circles are selected to be the local set. A index matrix is used to store the indices of the unknowns. The matrix structure shown in the figure has three layers. First layer is the \emph{absolute} layer which keeps the indices of the global set. The second layer is the \emph{relative} layer which stores the relative indices of the unknowns of $\mathcal{S}_L$. Last layer, depicted as the \emph{flag} layer is actually a layer of reference to construct the local set. At the beginning, all of the entries of this layer is 1 meaning that all of the variables are part of the local set. Depending on the set rule, either 0 or 1 assigned for each entry. While scanning the domains to construct the set, point with 0 flag values are ignored. Examining the figure, it can be seen that all rows are not fully covered, even $j=4$ has no points in the set. In this model, size$\left(\mathcal{S}_L\right)=11$ and size$\left(\mathcal{S}_G\right)=25$. Therefore, $\mathbf{f}$ should return a vector with 11 elements only.

\section{Implementation}\label{algorithm}
\begin{figure}[htbp]
\centering
\includegraphics[scale=0.5]{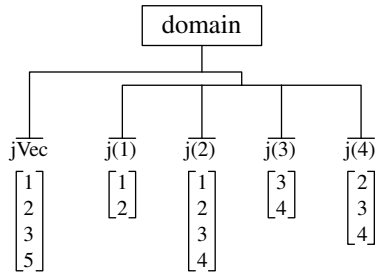}
\caption {Set tree}
\label{f:setTree}
\end{figure}

When a new discretized problem is implemented into the solver, the model is presented for all of the unknowns. Adding special cases with \texttt{if} or \texttt{switch} clauses will slow down the computations. On the other hand, set idea seems to work only with if statements where the indices are checked over the \emph{flag} layer. In this study, another approach is employed to evaluate $\mathbf{f}$ considering the local functions, $f_i$'s in the set. In order to understand the implementation of the set idea, it is constructive to advance step by step. To start, the unknowns have to be organized. The inventory is presented in Table \ref{t:set1}. $i$ indices are given for each $j$ with their respective absolute and relative indices. Absolute index is for the identification of the node on $\mathcal{S}_G$ and relative index is for $\mathcal{S}_L$. One imminent outcome is that $i$'s are actually members of $j$'s. For a given row, the columns can be stored in vectors with varying lengths. The respective tree structure is given in Figure \ref{f:setTree}.
\begin{table}[htbp]
\caption{Index inventory of Figure \ref{f:simpleGridSet}}
\begin{center}
\begin{tabular}{|c|c|c|c|}
  \hline
  $j$ & $i$'s of $j$ & absolute indices & relative indices\\  \hline
  1 & 1, 2 & 1, 2 & 1, 2\\ \hline
  2 & 1, 2, 3, 4 & 6, 7, 8, 9 & 3, 4, 5, 6\\ \hline
  3 & 3, 4 & 13, 14 & 10, 11 \\ \hline
  5 & 2, 3, 4 & 22, 23, 24 & 12, 13, 14 \\
  \hline
\end{tabular}
\end{center}
\label{t:set1}
\end{table}
As observed from \ref{f:setTree}, for each $j$, an $i$ vector is stored. Hence, instead of looping over $i$ and $j$'s, the cycles will be based on the vectors. This model is demonstrated in Table \ref{simpleFIM}. In this code segment, \texttt{jVec} is a vector that stores the indices of the $j$ that belong to the set. \texttt{j}'s on the other hand are derived types that keeps the indices of the $i$'s per $j$ in $iVec$. When Figure \ref{f:setTree} is examined, it can be seen that the lengths of the \texttt{iVec} do differ. This is considered algorithmically while forming the local set. Local set, $\mathcal{S}_L$, actually refers to these vectors that forms the tree-structure.
\begin{table}[htbp]
\caption{Classic loop vs. Set loop}
\centering
\begin{tabular}{| m{0.50\paperwidth} |}
\hline
\begin{footnotesize}
\begin{verbatim}


! classic loop style
do j=2,n-1
    do i=2,n-1
        ...
    end do
end do

! set loop style
do myJ=1,size(domain%jVec) ! access to jVec
    j=domain%jVec(myJ)
    do myI=1,size(domain%j(myJ)%iVec)
        i=domain%j(myJ)%iVec(myI) ! access to iVec of j(mjJ)
        ...
    end do
end do
\end{verbatim}
\end{footnotesize}
\\ \hline
\end{tabular}
\label{simpleFIM}
\end{table}
With the use of this idea, no if statement will be necessary. While solving Equation \ref{eq:jacobian}, the local set will be fixed and the computations will be based on Table \ref{simpleFIM}. Note that, when the linear solver is used, it will operate on a smaller $\Delta \mathbf{x}$ whose size is 11 for this example model.

While using the subsets, the solution vector $\mathbf{x}$ is not regenerated. Updates on the solution vector are applied regardless of the use sets. The key point is to use a vector, which is called \texttt{setVec}, that stores the absolute indices of the unknowns in $\mathcal{S}_L$.
The vector for the example is \texttt{setVec}=$[1,2,6,7,8,9,13,14,22,23,24]$. Thus, when the update is applied onto the solution following assignment in Equation \ref{eq:set} is used. When the solver works on the global set, then \texttt{setVec}=$1:25$ so the original system is recovered.
\begin{equation}
\mathbf{x}(\texttt{setVec})=\mathbf{x}(\texttt{setVec})+\lambda\Delta \mathbf{x}\label{eq:set}
\end{equation}

Now, the algorithm can be presented as in Table \ref{setalg}. At step 1, Newton's Method is applied on $\mathcal{S}_G$. By default, only one iteration is performed to calculate the update, $\Delta \mathbf{x}$ and to use a more realistic $\mathbf{f}$. Later in the second phase, the unknowns are tested against the set rules defined in section \ref{set_idea}. New subset is solved in step 4 and the update is applied on $\mathbf{x}$ using \texttt{setVec} in step 5. Last step is to check whether the solution satisfies Equation \ref{eq:nonlinear}.

\begin{table}[htbp]
\caption{Set Algorithm}
\centering
\begin{tabular}{| m{0.50\paperwidth} |}
\hline
\\
\begin{enumerate}
  \item Perform a computation on $\mathcal{S}_G$
  \item Apply the set rule and modify the \emph{flag} layer.
  \item Form $\mathcal{S}_L$ and \texttt{setVec}
  \item Solve $\mathcal{S}_L$
  \item Update $\mathbf{x}$ on $\mathcal{S}_L$ and test $\mathbf{f}$ on $\mathcal{S}_G$
  \item If converged stop; else goto 1
\end{enumerate}
\\ \hline
\end{tabular}
\label{setalg}
\end{table}

In step 4, the problem is solved until the solution is achieved for the local set. A variant of the algorithm above can be used at which one avoids the accurate solution of the local set and performs just one Newton iteration. This algorithm actually performs just a single series of Newton Method where the size of the problem varies from one step to another step. This variant is also tested in this study. Before completing this section, it should be note that the set idea resembles the block-Gauss Seidel method. What is different is that the elements of the block have a diverse pattern on the domain. Also note that, an improvement on the idea is to use the sets as the Newton's Method proceeds.

\section{Results and Discussion}\label{results}
For the solution of systems of nonlinear equations, Newton-GMRES is
used. Inexact Newton Method \cite{dembo_1982a} is applied with
adaptive forcing parameter \cite{eisenstat_1996a}. Matrix-vector
products in GMRES are performed using the matrix-free methodology as
in Equation \ref{eq:nk1}. Selection of the perturbation parameter
$\varepsilon$ is based on the idea presented in \cite{brown_1990a}.
Combination of absolute and relative tolerances are used to define
the nonlinear convergence tolerance as suggested in
\cite{kelley_1995a} with $\tau_{abs}=\tau_{rel}=1E-8$. Line search
backtracking is used to damp the updates \cite{dennis_1983a}.
Maximum number of nonlinear iterations is 20.
\begin{equation}
\mathbf{Jv}\approx \frac{\mathbf{f}(\mathbf{x}+\varepsilon \mathbf{v})-\mathbf{f}(\mathbf{x})}{\varepsilon}\label{eq:nk1}
\end{equation}
\subsection{Test Problem}
\begin{figure}[htbp]
\centering
\includegraphics[scale=0.5]{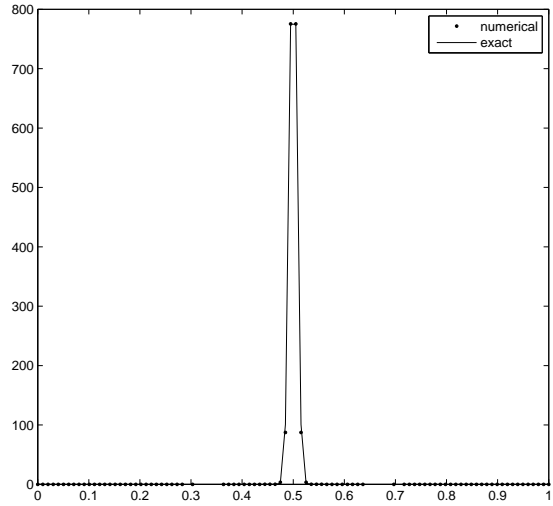}
\caption {Solution of the test problem}
\label{f:exact}
\end{figure}

\begin{figure}[htbp]
\centering
\includegraphics[scale=0.5]{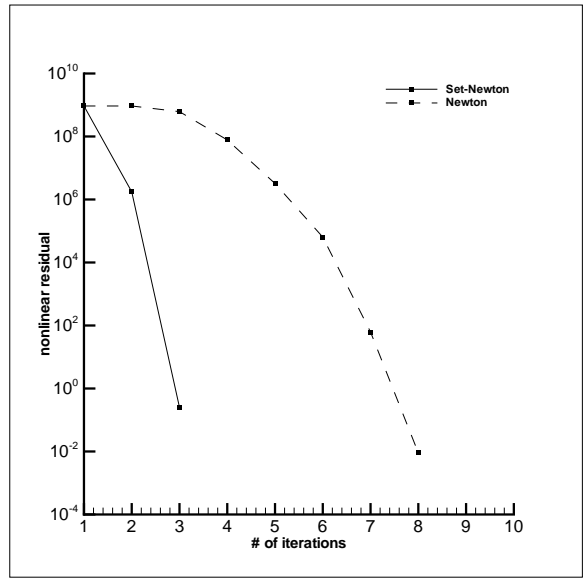}
\caption {Comparison of the set idea with Newton's Method}
\label{f:compare1}
\end{figure}

Test problem is taken from \cite{lanzkron_1996a} which reads as in Equation \ref{eq:test1}\footnote{The authors believe there are typos in the equation given in the reference. The exact solution does not satisfy the ODE unless it is corrected as in Equation \ref{eq:test1}} with boundary conditions $u(0)=u(1)=0$. The exact solution is given in Equation \ref{eq:test1e}.
\begin{equation}
 - u'' + u^3  + \left( {4 \times 10^8 \left( {x - 0.5} \right)^2  - 2 \times 10^4 } \right)u - 10^9 e^{ - 3\left( {\frac{{x - 0.5}}
{{0.01}}} \right)^2 }  = 0
\label{eq:test1}
\end{equation}
\begin{equation}
u(x)=10^3e^{-\left(\frac{x-0.5}{0.01}\right)^2}
\label{eq:test1e}
\end{equation}
The solution of the equation has a peak around $x=0.5$ which complicates the solution. In \cite{lanzkron_1996a}, the set is determined before solving the problem: For a 100 node grid, the initial set is fixed as nodes 49-52. In current study, these points are automatically determined using set rules. The solution of the problem is given in Figure \ref{f:exact}.
In Figure \ref{f:compare1}, Newton's Method is compared with the
solution based on the set idea. Excluding the initial norms,
Newton's Method converges in 7 steps whereas the set idea uses only
2 global iterations. In the sub solves of the set approach (i.e.
step 4 in Table \ref{setalg}) 6 and 3 Newton steps are performed,
respectively. Newton's Method required 34 total linear iterations
where the set idea needs 33 linear iterations. Although, set method
is 1 linear iteration shy, more linear iterations per nonlinear
iteration is needed compared to Newton's method (16.5 vs. 4.9). In
the first step of the set idea, the local set is identical to those
points used in \cite{lanzkron_1996a}, only 4 points (49:52) are kept
in $\mathcal{S}_L$. In the second iteration, however, the local set
is expanded to 12 points (45:56). In both cases, the local set is
around $x=0.5$ which makes sense because of the spike in the
problem.
This comparison is based by using average of the nonlinear residuals
with $\alpha=0.01$. When $\alpha=1$, then set still converges in two
steps but requires 35 total linear iterations. Use of RMS of
$\mathbf{f}$ performs the same as the average values. Use of the
step size criterion did not result a reduction in the set and the
computations are performed as Newton's Method. When the number of
unknowns is increased, it can be observed that the set idea requires
more linear iterations, however, the computations are to be
performed in less amount of time. In Table \ref{t:compare1}, the
methods are compared on different grid sizes where iters denotes the
nonlinear iterations and the numbers in the parenthesis are total
number of linear steps. Set size shows the size of the local set at
each set iteration. The sets are created using the average values
with $\alpha=0.001$. The efficiency of the proposed idea is apparent
over the Newton's method when computational times are considered.
The results in Table resembles \ref{t:compare1} the results in
\cite{lanzkron_1996a} but speedup over the Newton's Method is not
constant as in the reference.
\begin{table}[htbp]
\caption{Comparison in various grid sizes}
\begin{center}
\begin{tabular}{|c|c|c|c|c|c|}
  \hline
  grid size& Newton Time & Newton Iters & Set Time & Set Iters & Set size
\\  \hline
  500 & 0.4 s & 11(105) & 0.25 s &  3(137)& 20,36,26  \\ \hline
  1000 & 0.92 s& 10(159) & 0.51 s & 3(293)& 40,68,59 \\ \hline
  5000 & 111.17 s& 13(1422) & 17.38 s &4(2101) & 196,311,287,498 \\
  \hline
\end{tabular}
\end{center}
\label{t:compare1}
\end{table}

\section{Conclusion}
In this study, an idea is presented to reduce the solution set of a nonlinear system. The restriction of the set can be carried out by using different rules that are based on the nonlinear residual as well as the step size. A new approach is demonstrated to keep track of the unknowns on the local sets and to evaluate the nonlinear system. The idea is tried on a test problem and it is observed that use of local sets improves the solution.

This idea can be tested on harder problems in two dimensional geometries and with multi-dof systems. The variant of the algorithm can also be used to examine the effectiveness of the set idea. The method presented here is used only with finite differences. It can be extended to finite volumes and finite elements as well as to unstructured grids. A theoretical study was out of the scope of this paper, however, a numerical analysis could improve the selection of the sets and the tolerances.

\bibliographystyle{amsplain}

\bibliography{serne}

\end{document}